# A new transport phenomenon in nanostructures: A mesoscopic analog of the Braess paradox encountered in road networks.

Marco Pala[1], Hermann Sellier[2], Benoit Hackens[3], Frederico Martins[3], Vincent Bayot[2,3], Serge Huant[2,$]

1 : IMEP-LAHC, Grenoble INP, Minatec, BP 257, F-38016 Grenoble

2 : Institut Néel, CNRS & Université Joseph Fourier, BP 166, F-38042 Grenoble

3 : IMCN/NAPS, UCLouvain, 2 chemin du cyclotron, B-1348 Louvain-la-Neuve

$ Corresponding author

## Abstract

The Braess paradox, known for traffic and other classical networks, lies in the fact that adding a new route to a congested network in an attempt to relieve congestion can counter-intuitively degrade the overall network performance. Recently, we have extended the concept of Braess paradox to semiconductor mesoscopic networks, whose transport properties are governed by quantum physics. In this paper, we demonstrate theoretically that, alike in classical systems, congestion plays a key role in the occurrence of a Braess paradox in mesoscopic networks.

## Introduction

Adding a new road to a congested road network can paradoxically lead to a deterioration of the overall traffic situation, i.e. longer trip times for individual road users. Or, in reverse, blocking certain streets in a complex road network can surprisingly reduce congestion [1]. This counter-intuitive behavior has been known as the Braess paradox [2]. Later extended to networks in classical physics such as electrical or mechanical networks [3,4], this paradox lies in the fact that adding extra capacity to a congested network can counter-intuitively degrade its overall performance.

Known so far in classical networks only, we have recently extended the concept of the Braess paradox to the mesoscopic world [5]. By combining quantum simulations of a model system and scanning-gate microscopy [6-10], we have discovered that an analog of the Braess paradox can occur in mesoscopic electron networks, where transport is governed by quantum mechanics. To explore the possibility of a mesoscopic Braess paradox, we had set up a simple two-path network in form of a hollow rectangular corral connected to source and drain via two openings, with dimensions such that the embedded two-dimensional electron gas (2DEG) is in the ballistic and coherent regimes of electron transport at 4.2K. The short wires in the initial corral, Fig. 1(a), were narrower than the long wires in order to behave as congested constrictions for propagating electrons (see below). Branching out this basic network by adding a central wire as shown in Fig. 1(a) opens an additional path to the electrons. Then, we have used scanning gate microscopy [6-10] to partially block by local gate effects the electron transmission through this additional path. Doing so should intuitively result in a decreased total current transmitted through the device since one electron path partly looses efficiency, but we counter-intuitively found, both numerically and experimentally, that it is exactly the opposite behavior that can actually take place [5].

A key ingredient in the occurrence of classical Braess paradoxes is the network congestion. Our previous work was made on a congested mesoscopic network and it indeed exhibited a marked paradoxical behavior. In this letter, we study numerically in more detail the effect of congestion by simulating three rectangular corrals of different dimensions, that is, different degrees of congestion. We show that releasing congestion considerably relaxes the paradoxical behavior. Simulations of the spatial distribution of the current density inside the networks for different positions of the local gate help to interpret our predictions in terms of current redistribution inside the network.

## Theoretical details

The three simulated networks are shown in Fig. 1(a-c). The narrowest network in Fig. 1(a) is nearly identical to that simulated in our previous work [5], apart from slightly larger openings (320 nm instead of 300 nm). Its dimensions are chosen such that the electron flow is congested. Indeed, in a system where electrons can be backscattered solely by the walls defining the structure geometry, a sufficient condition to reach congestion is obtained when the number of conducting modes allowed by internal constrictions is smaller than the number of conducting modes in the external openings, which implies $2W < W_0$, where $W$ and $W_0$ denote the widths of the lateral arms (both of the same width) and of the external openings (of equal widths too), respectively. In turn, increasing $W$ such that $2W > W_0$ as shown in Fig. 1(b) progressively relaxes congestion since all conducting modes injected by the openings can be admitted in the lateral arms. Starting from the network of Fig. 1(b), we will further relax congestion by increasing the widths $L$ of the horizontal long arms, as shown in Fig. 1(c).

The transport properties of these structures are simulated within an exact numerical approach based on the Keldysh Green's function formalism. A thermal average is performed around the Fermi energy $E_F$ at the temperature $T$= 4.2 K. We adopt a mesh size of $\Delta x = \Delta y$= 2.5 nm. The Green's function of the system is computed in the real-space representation that allows us to take into account all possible conducting and evanescent modes. Moreover, in order to reduce the computational time and memory requirements we exploit a recursive algorithm, which is based on the Dyson equation [5,8].

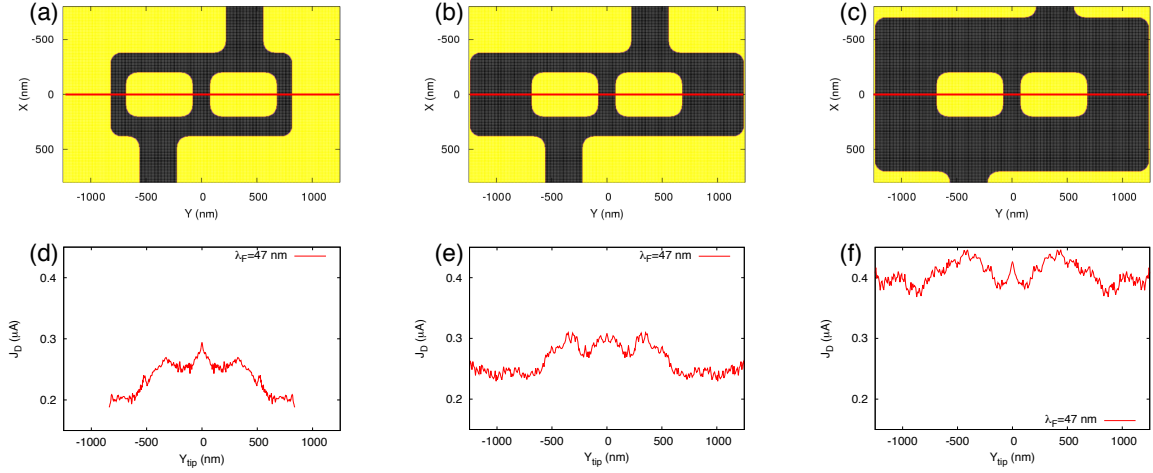

**Figure 1. Evidence for the key role of network congestion in the occurrence of the mesoscopic Braess paradox.** (a-c) depict the network geometries. All networks have a central (additional) branch of 160 nm width and 320 nm wide openings. In (a): $W$= 140 nm and $L$= 180 nm. In (b): $W$= 560 nm and $L$= 180 nm. In (c): $W$= 560 nm, $L$= 500 nm. (d-f) depict the current transmitted through the networks as function of the tip position scanned along the median lines of the networks (red lines in (a-c)). The source-drain voltage applied to the networks is $V_{ds}$= 1 mV and the potential applied to the tip is – 1V (see text). Fermi wavelength: $\lambda_F$= 47 nm, $T$= 4.2 K.

In this framework the current densities along the *x*-axis (transport direction) and the *y*-axis (transverse direction) between two adjacent nodes read:

$$J_{i,i+1;k,k} = -\frac{4e}{h} \int d\omega \, \text{Re}\left[H_{i,i+1;k,k} G^{<}_{i+1,i;k,k}(\omega)\right]$$

$$J_{i,i;k,k+1} = -\frac{4e}{h} \int d\omega \, \text{Re}\left[H_{i,i;k+1,k} G^{<}_{i,i;k+1,k}(\omega)\right]$$

where $H_{i,i';k,k'}$ represents the Hamiltonian discretized on the local basis and $G^{<}_{i,i';k,k'}(\omega)$ is the lesser-than Green's function in the real space representation and energy domain.

The tip-induced potential is simulated by considering a point-like gate voltage of -1 V placed at 100 nm above the 2DEG, which corresponds to a lateral extension of ≈ 400 nm for the tip-induced potential perturbation at the 2DEG level.

## The key role of congestion in the network

Figs. 1(d-f) show the current flowing through the structures depicted in Figs. 1(a-c), respectively, as a function of the tip position scanned along the median lines (red lines). Fig. 1(d) shows the occurrence of an analog of the classical Braess paradox in a congested mesoscopic network as a distinctive current peak centered at $Y_{tip}$=0 nm. When the tip induced potential closes the central wire connecting the two openings in Fig. 1(a) the current is counter-intuitively increased. However, Figs. 1(e,f) show that as soon as the condition for congestion is relaxed, allowing a larger number of conducting channels to propagate in the region inside the structure, the paradox disappears and the total current exhibits a maximum when the tip is placed over the two antidots.

In order to microscopically study this behavior, we have simulated in Fig. 2 the spatial distribution of the absolute value of the current $|J|$ inside the three structures for $Y_{tip}$=0 nm and $Y_{tip}$=-400 nm. When comparing Fig. 2(a) and Fig. 2(d) for the congested structure, we can notice that the opening of a third central wire connecting the contacts has a twofold effect. The first consequence is to create a direct connection between source and drain, which should positively contribute to the total current flowing through the system. The second one is to generate alternative paths that trap electrons in the central region and should promote a longer stay inside the network. We believe that this second effect is the one responsible for the decrease of the total current as far as the third wire is opened. The comparison of Fig. 2(a) and (d) is indeed very instructive, and in particular the behavior of current through the right path which, paradoxically, decreases while the depleting tip moves away. This behavior clearly indicates that the current contribution of trapped electrons around the right anti-dot compensates partially the initial current. This effect is only partly replicated in the networks of Fig. 1(b) and 1(c), whose current redistributions are shown in Figs. 2 (b,e) and Figs. 2 (c,f), respectively. In these cases the re-opening of the third wire obtained by placing the tip over the antidot induces a number of new internal paths, which is small compared to the large number of semiclassical trajectories already present in the lateral arms. Therefore, the closing of the central path implies only a small current increase in Figs. 1(e) and 1(f) around the position $Y_{tip}$=0 nm, which is not sufficient to overcome the current at $Y_{tip}$= -400 nm.

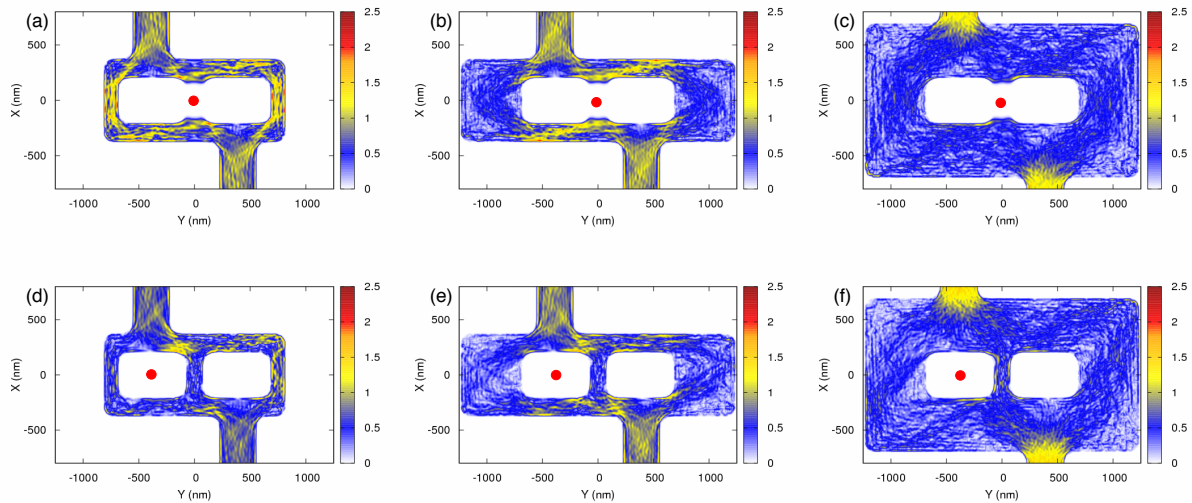

**Figure 2. Current redistribution in the mesoscopic networks.** All figures depict contour plots of the spatial distribution of the current density. (a-c): the tip, marked by a red dot, is positioned above the middle of the networks, that is, above the center of the additional arm. (d-f): the depleting tip is positioned above the center of the left hand side antidot. Fermi wavelength $\lambda_F$= 47 nm, $V_{ds}$= 1 mV, $T$= 4.2 K.

## The robustness of the paradox

Finally, in order to test the robustness of our results we simulated the non congested structure of Fig. 1(c) at different Fermi wavelengths ($\lambda_F$= 57, 47 and 38 nm). This is shown in Fig. 3. The behavior of the three curves is qualitatively very similar: they present two regions of maximum current when the gated tip is placed over the two antidots, allowing the passage of electrons through the central path, but they also show a local increase in current around $Y_{tip}$= 0, when the tip closes the central path. This is a signature that the mechanism responsible for the occurrence of the paradox in the congested structure of Fig.1 (a), even if still present, is not predominant with respect to the direct coupling between the two contacts provided by the third wire.

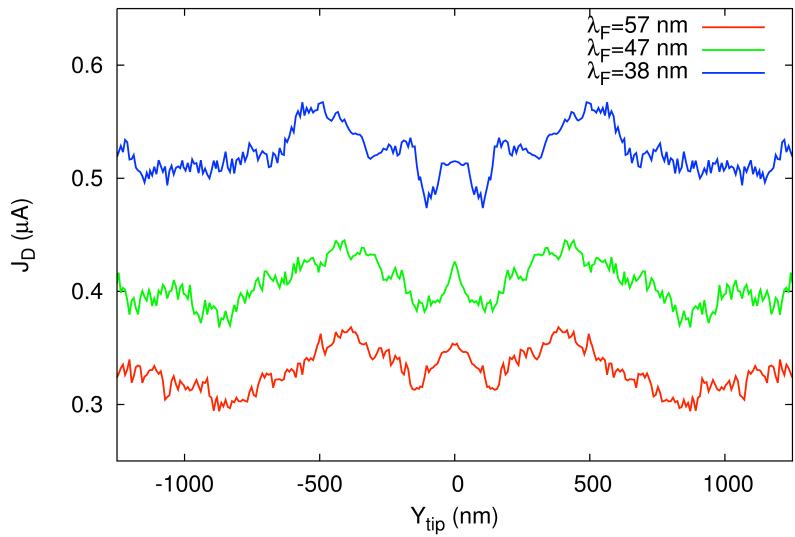

**Figure 3. Robustness of the results.** The current transmitted through the network of Fig.1(c) as a function of the tip position scanned along the median line (red lines in Fig.1(c)) for three different Fermi wavelengths $\lambda_F$= 57,47 and 38 nm. The potential applied to the tip is -1V, $V_{ds}$= 1 mV, $T$= 4.2 K.

## Conclusions

In this letter, we have studied the geometric conditions of mesoscopic networks for the occurrence of a quantum analog of the Braess paradox, known previously for classical systems only. By analyzing the spatial distribution of current density in different structures we have shown that congested structures are the most suitable geometries to the occurrence of such a counter-intuitive phenomenon. This is reminiscent to what is known for the classical paradoxes, in particular for the historic road-network Braess paradox.

## Competing interests

The authors declare that they have no competing interests.

## Authors' contributions

MP performed all of the simulations. SH initiated the work and presented the talk at ICSNN2012. MP and SH wrote the paper. All authors animated the discussions on the Braess paradox, extensively discussed the results, and proofread the article.

## Acknowledgments

This work has been supported by the French Agence Nationale de la Recherche ("MICATEC" project), the FRFC Grant No. 2.4.546.08.F and FNRS Grant No. 1.5.044.07.F, and by the Belgian Science Policy (Program IAP-6/42). V. B. acknowledges support from the Grenoble Nanosciences Foundation ("Scanning-Gate Nanoelectronics" project).